\documentclass[11pt]{article}
\usepackage{graphicx,hyperref}
\usepackage{
bm}

\usepackage{amsmath,amssymb}
\usepackage{makeidx}
\makeindex
\usepackage{ascmac}
\usepackage{pb-diagram}
\usepackage{here}
\usepackage{esint}
\title{Extended Series of   
Correlation
 Inequalities in Quantum Systems
 }

\author{Chigak Itoi$^1$, 
Hiroto Ishimori $^1$, Kota Sato$^1$ and Yoshinori Sakamoto$^2$\\
\\
$^1$Department of Physics, GS $\&$ CST, Nihon University, \\
Kanda-Surugadai, Chiyoda, Tokyo 101-8308, Japan
 \\
$^2$ Laboratory of Physics, CST, Nihon University, \\
Narashinodai, Funabashi-city, Chiba 274-8501, Japan} 

\begin{document}
\maketitle
\begin{abstract}{
A systematic derivation provides extended series of correlation inequalities in quantum systems. 
Each order in truncated Taylor expansion of the spectral representation
 for  the Duhamel correlation  function  gives its  lower and upper  bounds.  
The obtained bound on the Duhamel function and the square root interpolation method enable us to derive a variational solution of specific free energy in the transverse field Sherrington-Kirkpatrick model. 
}\end{abstract}
\section{Introduction}
Spectral representations of physical observables are known to be useful to study quantum statistical systems.  
Famous correlation inequalities  are obtained in these representations, 
 such as the Bogoliubov inequality, the Harris 
inequality and the Falk-Bruch inequality \cite{B,BT,FB,H,R,S}.
These inequalities provide bounds on physical observables, which enable us to prove rigorous theorems on these observables in quantum systems. 
The Mermin-Wagner theorem \cite{MW} is 
proven using the Bogoliubov correlation inequality. This theorem claims that the spontaneous symmetry breaking of continuous symmetry
cannot occur at any finite temperature in two or one dimensions.  
Recently, Leschke, Manai, Ruder and Warzel have proven the  non-zero variance of the overlap operator in the transverse Sherrington-Kirkpatrick (SK) model \cite{W}, using the Falk-Bruch inequality \cite{FB,R}. Many researchers who study spin glass systems appreciate their result. 
Alternatively, their result can be proven more easily by the Harris inequality \cite{BT,H} or other extended inequalities instead of the Falk-Bruch inequality. 

In the present paper, several  correlation  inequalities 
are obtained systematically in terms of spectral representations of operators. 
The present paper is organized as follows. In section 2, definitions of several complex valued functions and main results of new correlation inequalities are provided. In section 3, the main results are  proven in terms of contour integrations
of the spectral representation for operators. In section 4, obtained correlation 
inequalities are applied to the transverse field SK model.
We extend the square root interpolation for a variational solution of the replica symmetric
specific free energy  given by Guerra and Talagrand \cite{G1,T}
to the quantum mechanically  perturbed model.

\section{Definitions and Main Results}
Consider a quantum system with a Hamiltonian $H$.
The Duhamel function for bounded linear operators $A, B$ is defined by
\begin{equation}
 (A,B) =\int_0^1 dt \langle e^{\beta t H} A e^{-\beta tH} B \rangle,
 \end{equation}
 which is important to represent a susceptibility of quantum spin systems.  

To express the main theorem, 
consider  functions  $f : \mathbb C \to \mathbb C$, $g : \mathbb C \to \mathbb C$ and $h : \mathbb C \to \mathbb C$ defined by
\begin{equation}
f(z) :=z\coth z, \   \ g(z):=\frac{1}{f(z)}=\frac{ \tanh z}{ z}, \ \ h(z):= \frac{z}{\log
\frac{1+z}{1-z}}
.
\label{fgh}
\end{equation}
 Note that $f(-z) = f(z)$, $g(-z) = g(z)$ and $h(-z) = h(z)$. 

Define  $k$-th differential coefficient of the following operator is defined by
\begin{equation}
C_A^k:= \Big( \frac{d^k}{dt^k}  e^{tH} A e^{-tH} \Big)_{t=0} = [H,  \cdots,[H, [H,A] ]\cdots ],
\label{Ck}
\end{equation}
for an arbitrary  positive integer $k$,  and define $C_A^0 := A$. 
\\
\paragraph{Theorem 1} {\it \label{MT}  Let $A$ be a bounded linear operator and $n$ be a positive even integer, such that $n/2$ is odd.
The difference between the expectation of
the anti-commutator and the Duhamel function is bounded by the following sequence with expectations of  double commutators:
\begin{equation}
\sum_{k=1}^{n/2+1} \Big(\frac{\beta}{2}\Big)^{2k-1}\frac{f^{(2k)}(0)}{2(2 k)!} \langle  [{C_A^{k-1}}^\dag, C_A^{k}] \rangle
 \leq \frac{1}{2} \langle \{A^\dag,A \}\rangle-  (A^\dag, A)  \leq 
 \sum_{k=1}^{n/2} \Big(\frac{\beta}{2}\Big)^{2k-1}\frac{f^{(2k)}(0)}{2(2 k)!}\langle  [{C_A^{k-1}}^\dag,  C_A^{k}] \rangle.
 \label{MTineq}
\end{equation}
}

\paragraph{Theorem 2}
{\it Let $A$ be a bounded linear operator and $n$ be a nonnegative even integer, such that $n/2$ is even.
The Duhamel function is bounded by the following sequence with  the expectation of
the anti-commutators:
\begin{equation}
\sum_{k=0}^{n/2+1} \Big(\frac{\beta}{2}\Big)^{2k}\frac{g^{(2k)}(0)}{2(2 k)!} \langle  \{ {C_A^k}^\dag, C_A^k \} \rangle
 \leq
  (A^\dag, A)  \leq 
 \sum_{k=0}^{n/2} \Big(\frac{\beta}{2}\Big)^{2k}\frac{g^{(2k)}(0)}{2(2 k)!}\langle  \{ {C_A^k}^\dag, C_A^k \} \rangle.
 \label{MT2ineq}
\end{equation}
}

\paragraph{Theorem 3}
{\it  Let $A$ be a bounded linear operator and $n$ be a nonnegative even integer, such that $n/2$ is even.
The following expectation of commutator is bounded by  sequences with  the expectation of
the anti-commutators:
\begin{equation}
\sum_{k=0} ^{n/2+1} \frac{g^{(2k)} (0)}{(2k)!} \Big( \frac{\beta}{2}\Big)^{2k+1} \langle \{ {C_A^{k+1}}^\dag,C_A^{k+1} \} \rangle
\leq \langle[A^\dag, [H, A] ]\rangle
\leq \sum_{k=0} ^{n/2} \frac{g^{(2k)} (0)}{(2k)!} \Big( \frac{\beta}{2}\Big)^{2k+1} \langle \{ {C_A^{k+1}}^\dag,C_A^{k+1} \} \rangle.
\label{MT3ineq}
\end{equation}}

\paragraph{Theorem 4}
{\it Let $A$ be a bounded linear operator and $n$ be a nonnegative even integer, such that $n/2$ is even.
The Duhamel function is bounded by the following sequence with  the expectation of
the commutator and  anti-commutator:
\begin{equation}
(A^\dag,A) \leq \langle \{A^\dag,A\}\rangle \sum_{k=0}^{n/2} \frac{h^{(2k)}(0)}{(2k)!} \Big(\frac{\langle [A^\dag, A]\rangle}{\langle \{A^\dag,A\}\rangle}\Big)^{2k}.
\end{equation}
}
 For $n=2$,  Theorem 1 implies
\begin{equation}
\frac{\beta}{12} \langle  [A^\dag, [H, A]] \rangle -\frac{\beta^3}{720}   \langle  [[H,A]^\dag, [H,[H, A]]] \rangle
\leq  \frac{1}{2} \langle \{A^\dag,A \}\rangle-  (A^\dag, A)\leq \frac{\beta}{12} \langle  [A^\dag, [H, A]] \rangle.
\label{Ex1}
\end{equation}
The upper bound is known as the Bogoliubov-Harris inequality \cite{ BT,H},  and the lower bound gives a new inequality. 

For $n=0$,  Theorem 2 implies
\begin{equation}
 \frac{1}{2}  \langle  \{ A^\dag, A\} \rangle-\frac{\beta^2}{24}   \langle  \{[H,A]^\dag, [H, A] \} \rangle
\leq (A^\dag, A)\leq  \frac{1}{2}  \langle  \{ A^\dag, A\} \rangle.
\label{Ex2}
\end{equation}
Note that  the upper bound 
is well-known, and the lower bound is a new one given by the left hand side.
Several inequalities given by Theorem 1 and  2 have been obtained by Brankov and Tonchev \cite{BT}. 

For $n=0$,  Theorem 3 implies
\begin{equation}
 \frac{\beta}{2}  \langle  \{[H,A]^\dag, [H, A] \} \rangle-\frac{\beta^3}{24}   \langle  \{[H,[H,A]]^\dag, [H,[H, A]] \} \rangle
\leq \langle [A^\dag,[H, A]] \rangle \leq  \frac{\beta}{2} \langle  \{[H,A]^\dag, [H, A] \} \rangle.
\label{Ex3}
\end{equation}
These are new inequalities. 

For $n=4$, Theorem 4 gives a new inequality
\begin{equation}
\frac{2(A^\dag,A)}{ \langle \{A^\dag,A\}\rangle} \leq  1- \frac{1}{3} \Big(\frac{\langle [A^\dag, A]\rangle}{\langle \{A^\dag,A\}\rangle}\Big)^2
-\frac{4}{45}\Big(\frac{\langle [A^\dag, A]\rangle}{\langle \{A^\dag,A\}\rangle}\Big)^4.
\end{equation}

\section{Proofs}
 Spectral representation is well-known as a useful method to represent correlation functions in quantum systems \cite{S}.
To define a spectral representation of correlation functions, define 
energy eigenstates $|\mu \rangle$ belonging to the energy eigenvalue $E_\mu$
\begin{equation}
H | \mu \rangle = E_\mu |\mu \rangle.
\end{equation}  
The partition function for inverse temperature $\beta$ is 
\begin{equation}
Z(\beta) := {\rm Tr} e^{-\beta H} = \sum_{\mu} e^{-\beta E_\mu}.
\end{equation}
Define spectral function $Q_{A,B}(\omega)$ of bounded  linear operators $A, B$ for $\omega \in \mathbb R$ by  
\begin{equation}
Q_{A,B}(\omega) := \frac{1}{Z(\beta)} \sum_{\mu,\nu} e^{-\beta E_\nu } \langle \nu | A| \mu \rangle\langle \mu | B| \nu \rangle (1+e^{-\beta \omega}) \delta (E_\mu-E_\nu-\omega).
\end{equation}
The function $Q_{A,B}(\omega)$ has the following properties. $Q_{A,B}(\omega)$ is bilinear in $A$ and $B$. The complex conjugate is given by $Q_{A,B}(\omega)^* = Q_{B^\dag, A^\dag}(\omega)$.  $Q_{A,B} (-\omega) =Q_{B,A}(\omega)$. $Q_{A^\dag, A}(\omega) \geq 0$,  and $Q_{A^\dag,A}(\omega)=0$ implies 
$\langle A \rangle =0.$. 
Spectral representations for   
several  correlation functions are given  in the following.

\paragraph{Lemma 1} {\it The spectral representation of the expectation of  the anti-commutator between $A, B$ is given by
\begin{equation} \langle \{A, B\} \rangle = \langle (AB+BA) \rangle = \int_{-\infty}^\infty  d\omega Q_{A,B}(\omega).
 \end{equation}
 Proof.} The right hand side is   
\begin{eqnarray}
RHS&=&\int_{-\infty}^\infty d\omega  \frac{1}{Z(\beta)}
\sum_{\nu,\mu } e^{-\beta E_\nu } \langle \nu | A| \mu \rangle \langle \mu| B|\nu \rangle (1+e^{-\beta \omega})\delta (E_\mu-E_\nu -\omega)\nonumber \\
&=&  \frac{1}{Z(\beta)}\sum_{\nu,\mu } e^{-\beta E_\nu } \langle \nu | A| \mu \rangle \langle \mu| B|\nu \rangle(1 +  e^{ -\beta(E_\mu-E_\nu)})\nonumber\\
&=&  \frac{1}{Z(\beta)}\sum_{\nu,\mu }   \langle \nu | A| \mu \rangle \langle \mu| B|\nu \rangle(e^{-\beta E_\nu } +  e^{ -\beta E_\mu})\nonumber\\
&=&  \frac{1}{Z(\beta)}\sum_{\nu,\mu } (  \langle \nu | A| \mu \rangle \langle \mu| B|\nu \rangle e^{-\beta E_\nu } +  
 \langle \mu | A| \nu \rangle \langle \nu| B|\mu \rangle e^{ -\beta E_\nu})\nonumber\\
&=&  \frac{1}{Z(\beta)}\sum_{\nu,\mu } ( \langle \nu | A| \mu \rangle \langle \mu| B|\nu \rangle +
\langle \nu | B|\mu \rangle \langle \mu|A |\nu\rangle )e^{-\beta E_\nu }.
 \end{eqnarray}
Since  $1= \sum_{\mu } | \mu \rangle\langle \mu| $,  the left hand side is
\begin{eqnarray}
LHS&=& \frac{1}{Z(\beta)} {\rm Tr} (AB+BA) e^{-\beta H} = \frac{1}{Z(\beta)}\sum_{\nu }  \langle \nu | (AB+BA) e^{-\beta H}|\nu\rangle\\
&=&  \frac{1}{Z(\beta)}\sum_{\nu,\mu } ( \langle \nu | A| \mu \rangle \langle \mu| B|\nu \rangle +
\langle \nu | B|\mu \rangle \langle \mu|A |\nu\rangle )e^{-\beta E_\nu }.
 \end{eqnarray}
This is identical to the right hand side. $\Box$\\

\paragraph{Lemma 2} {\it The spectral representation of the expectation of the commutator between $A, B$ is given by
\begin{equation} \langle [A, B] \rangle = \int_{-\infty}^\infty  d\omega\tanh \frac{\beta \omega}{2} Q_{A,B}(\omega) 
 \end{equation}
Proof.} The right hand side is   
 \begin{eqnarray}
RHS&=&\int_{-\infty}^\infty d\omega \tanh \frac{\beta \omega}{2} \frac{1}{Z(\beta)}
\sum_{\nu,\mu } e^{-\beta E_\nu } \langle \nu | A| \mu \rangle \langle \mu| B|\nu \rangle (1+e^{-\beta \omega})\delta (E_\mu-E_\nu -\omega)\nonumber \\
&=&  \frac{1}{Z(\beta)}\sum_{\nu,\mu }  \tanh \frac{\beta (E_\mu-E_\nu)}{2} e^{-\beta E_\nu } \langle \nu | A| \mu \rangle \langle \mu| B|\nu \rangle(1 +  e^{ -\beta(E_\mu-E_\nu)})\nonumber\\
&=&  \frac{1}{Z(\beta)}\sum_{\nu,\mu }   \langle \nu | A| \mu \rangle \langle \mu| B|\nu \rangle
 \frac{e^{\beta(E_\mu-E_\nu)/2} - e^{-\beta(E_\mu-E_\nu)/2}}{e^{\beta(E_\mu-E_\nu)/2} + e^{-\beta(E_\mu-E_\nu)/2}}
(e^{-\beta E_\nu } +  e^{ -\beta E_\mu})\nonumber\\
&=&  \frac{1}{Z(\beta)}\sum_{\nu,\mu }   \langle \nu | A| \mu \rangle \langle \mu| B|\nu \rangle
 \frac{e^{-\beta E_\nu} - e^{-\beta E_\mu}}{e^{-\beta E_\mu} + e^{-\beta E_\nu}}
(e^{-\beta E_\nu } +  e^{ -\beta E_\mu})\nonumber\\
&=&  \frac{1}{Z(\beta)}\sum_{\nu,\mu }  \langle \nu | A| \mu \rangle \langle \mu| B|\nu \rangle 
( e^{-\beta E_\nu }-e^{-\beta E_\mu }) \nonumber\\
&=&  \frac{1}{Z(\beta)}\sum_{\nu,\mu } ( \langle \nu | A| \mu \rangle \langle \mu| B|\nu \rangle -
\langle \nu | B|\mu \rangle \langle \mu|A |\nu\rangle )e^{-\beta E_\nu }.
 \end{eqnarray}
 Assume resolution of unity  $1= \sum_{\mu } | \mu \rangle\langle \mu|$, then the left hand side is
\begin{eqnarray}
LHS&=& \frac{1}{Z(\beta)} {\rm Tr} [A,B] e^{-\beta H} = \frac{1}{Z(\beta)}\sum_{\nu }  \langle \nu | (AB-BA) e^{-\beta H}|\nu\rangle\nonumber\\
&=&  \frac{1}{Z(\beta)}\sum_{\nu,\mu } ( \langle \nu | A| \mu \rangle \langle \mu| B|\nu \rangle -
\langle \nu | B|\mu \rangle \langle \mu|A |\nu\rangle )e^{-\beta E_\nu }. 
 \end{eqnarray}
This is identical to the right hand side. $\Box$\\

\paragraph{Lemma 3}{\it The spectral representation of the expectation of the double commutator is given by  
\begin{equation}
  \langle [ A, [H ,B]] \rangle = \int_{-\infty}^\infty d\omega\omega \tanh \frac{\beta \omega}{2} Q_{A,B}(\omega).
\end{equation}
Proof.}  The right hand side is
 \begin{eqnarray}
RHS&=&\int_{-\infty}^\infty d\omega \omega \tanh \frac{\beta \omega}{2} \frac{1}{Z(\beta)}
\sum_{\nu,\mu } e^{-\beta E_\nu } \langle \nu | A| \mu \rangle \langle \mu| B|\nu \rangle (1+e^{-\beta \omega})\delta (E_\mu-E_\nu -\omega) \nonumber  \\
&=&  \frac{1}{Z(\beta)}\sum_{\nu,\mu }(E_\mu-E_\nu)  \tanh \frac{\beta (E_\mu-E_\nu)}{2} 
e^{-\beta E_\nu } \langle \nu | A| \mu \rangle \langle \mu| B|\nu \rangle(1 +  e^{ -\beta(E_\mu-E_\nu)})\nonumber\\
&=&  \frac{1}{Z(\beta)}\sum_{\nu,\mu }   \langle \nu | A| \mu \rangle \langle \mu| B|\nu \rangle (E_\mu-E_\nu)
 \frac{e^{\beta(E_\mu-E_\nu)/2} - e^{-\beta(E_\mu-E_\nu)/2}}{e^{\beta(E_\mu-E_\nu)/2} + e^{-\beta(E_\mu-E_\nu)/2}}
(e^{-\beta E_\nu } +  e^{ -\beta E_\mu})\nonumber\\
&=&  \frac{1}{Z(\beta)}\sum_{\nu,\mu }   \langle \nu | A| \mu \rangle \langle \mu| B|\nu \rangle (E_\mu-E_\nu)
 \frac{e^{-\beta E_\nu} - e^{-\beta E_\mu}}{e^{-\beta E_\mu} + e^{-\beta E_\nu}}
(e^{-\beta E_\nu } +  e^{ -\beta E_\mu})\nonumber\\
&=&  \frac{1}{Z(\beta)}\sum_{\nu,\mu }  \langle \nu | A| \mu \rangle \langle \mu| B|\nu \rangle (E_\mu-E_\nu)
( e^{-\beta E_\nu }-e^{-\beta E_\mu }) .
 \end{eqnarray}
 Since $1= \sum_{\mu } | \mu \rangle\langle \mu| $,  the left hand side is
\begin{eqnarray}
LHS&=& \frac{1}{Z(\beta)} {\rm Tr} [A,[H, B] ]e^{-\beta H} = \frac{1}{Z(\beta)}\sum_{\nu }  \langle \nu | [A (HB-BH)  -  (HB-BH)A ]e^{-\beta H}|\nu\rangle \nonumber \\
&=&  
 \frac{1}{Z(\beta)}\sum_{\nu }  \langle \nu | ( AHB- E_\nu A B - B A E_\nu +BHA) e^{-\beta E_\nu }|\nu\rangle\nonumber\\
 &=& \frac{1}{Z(\beta)}\sum_{\nu,\mu } (  \langle \nu | A| \mu \rangle   \langle \mu| B|\nu \rangle( E_\mu-E_\nu) -(E_\nu -E_\mu)
\langle \nu | B|\mu \rangle \langle \mu|A |\nu\rangle )e^{-\beta E_\nu }
\nonumber\\
&=&  \frac{1}{Z(\beta)}\sum_{\nu,\mu }  \langle \nu | A| \mu \rangle \langle \mu| B|\nu \rangle (E_\mu-E_\nu)
( e^{-\beta E_\nu }-e^{-\beta E_\mu }) .
 \end{eqnarray}
 This is identical to the right hand side. $\Box$\\
\paragraph{Lemma 4} {\it The Duhamel function for bounded linear operators $A, B$
\begin{equation}
 (A,B) = \int_{-\infty}^\infty \frac{d\omega}{\beta \omega} \tanh \frac{\beta \omega}{2} Q_{A,B}(\omega) 
\end{equation}
Proof.}  The right hand side is
 \begin{eqnarray}
RHS&=&\int_{-\infty}^\infty \frac{d\omega}{\beta \omega} \tanh \frac{\beta \omega}{2} \frac{1}{Z(\beta)}
\sum_{\nu,\mu } e^{-\beta E_\nu } \langle \nu | A| \mu \rangle \langle \mu| B|\nu \rangle (1+e^{-\beta \omega})\delta (E_\mu-E_\nu -\omega)\nonumber \\
&=&  \frac{1}{ Z(\beta)}\sum_{\nu,\mu }\frac{1}{\beta(E_\mu-E_\nu)}  \tanh \frac{\beta (E_\mu-E_\nu)}{2} 
e^{-\beta E_\nu } \langle \nu | A| \mu \rangle \langle \mu| B|\nu \rangle(1 +  e^{ -\beta(E_\mu-E_\nu)})\nonumber\\
&=&  \frac{1}{Z(\beta)}\sum_{\nu,\mu }   \langle \nu | A| \mu \rangle \langle \mu| B|\nu \rangle \frac{1}{\beta(E_\mu-E_\nu)}
 \frac{e^{\beta(E_\mu-E_\nu)/2} - e^{-\beta(E_\mu-E_\nu)/2}}{e^{\beta(E_\mu-E_\nu)/2} + e^{-\beta(E_\mu-E_\nu)/2}}
(e^{-\beta E_\nu } +  e^{ -\beta E_\mu})\nonumber\\
&=&  \frac{1}{Z(\beta)}\sum_{\nu,\mu }   \langle \nu | A| \mu \rangle \langle \mu| B|\nu \rangle \frac{1}{\beta(E_\mu-E_\nu)}
 \frac{e^{-\beta E_\nu} - e^{-\beta E_\mu}}{e^{-\beta E_\mu} + e^{-\beta E_\nu}}
(e^{-\beta E_\nu } +  e^{ -\beta E_\mu})\nonumber\\
&=&  \frac{1}{Z(\beta)}\sum_{\nu,\mu }  \langle \nu | A| \mu \rangle \langle \mu| B|\nu \rangle 
\frac{ e^{-\beta E_\nu }-e^{-\beta E_\mu }}{\beta(E_\mu-E_\nu)} .
 \end{eqnarray}
 Since $1= \sum_{\mu } | \mu \rangle\langle \mu| $, the left hand side is
\begin{eqnarray}
LHS&=& \int_0^1 dt \langle e^{\beta t H} A e^{-\beta tH} B \rangle\nonumber\\
&=&  \int_0^1 dt 
 \frac{1}{Z(\beta)}\sum_{\nu }  \langle \nu |e^{\beta t H} A e^{-\beta tH} B e^{-\beta E_\nu }|\nu\rangle\nonumber\\
 &=&\int_0^1 dt  \frac{1}{Z(\beta)}\sum_{\nu,\mu } \langle \nu |e^{\beta t E_\nu } A e^{-\beta t E_\mu}  | \mu \rangle \langle \mu| B e^{-\beta E_\nu }|\nu\rangle\nonumber\\
&=&  \frac{1}{Z(\beta)}\sum_{\nu,\mu }  \langle \nu | A| \mu \rangle \langle \mu| B|\nu \rangle e^{-\beta E_\nu}  \int_0^1 dt e^{t\beta (E_\nu-E_\mu)}\nonumber \\
&=&  \frac{1}{Z(\beta)}\sum_{\nu,\mu }  \langle \nu | A| \mu \rangle \langle \mu| B|\nu \rangle e^{-\beta E_\nu}
\frac{e^{ \beta (E_\nu-E_\mu)} -1}{\beta (E_\nu-E_\mu)}.
 \end{eqnarray}
This is identical to the right hand side. $\Box$\\

\paragraph{Lemma 5} {\it \label{HA} For an arbitrary bounded operators $A,B$ and for arbitrary positive integer $k$, 
the following identities are valid
\begin{equation}
 \omega^{k} Q_{A, B}(\omega) = Q_{A,  C_B^k}(\omega), \ \  \omega^{2k} Q_{A^\dag, A}(\omega) = Q_{{C_A^k}^\dag, C_A^k}(\omega). 
\end{equation} 
Proof.} 
For arbitrary bounded linear operators $A, B$, the following is valid
\begin{eqnarray}
Q_{A, [H,B]}(\omega) &=&\frac{1}{Z} \sum_{\mu,\nu} e^{-\beta E_\nu} \langle \nu | A | \mu\rangle   \langle \mu | [H,B] | \nu\rangle   (1+e^{-\beta \omega})
\delta(E_\mu-E_\nu -\omega) \nonumber \\
&=&\frac{1}{Z} \sum_{\mu,\nu} e^{-\beta E_\nu} \langle \nu | A | \mu\rangle   \langle \mu |(E_\mu-E_\nu) B| \nu\rangle   (1+e^{-\beta \omega})
\delta(E_\mu-E_\nu -\omega)\nonumber \\
&=&\frac{1}{Z} \sum_{\mu,\nu} e^{-\beta E_\nu} \langle \nu | A | \mu\rangle \omega  \langle \mu | B| \nu\rangle   (1+e^{-\beta \omega})
\delta(E_\mu-E_\nu -\omega)\nonumber
\\&=& \omega Q_{A,B}(\omega)
\end{eqnarray}
Therefore, the first identity is valid for $k=1$. Also, the identity for $k>1$ is obtained by successive use of the above identity.   
Since $H^\dag=H, $ and $[H,A]^\dag= [A^\dag, H]$, 
\begin{eqnarray}
Q_{[H,A]^\dag, [H,A]}(\omega) &=&\frac{1}{Z} \sum_{\mu,\nu} e^{-\beta E_\nu} \langle \nu |[A^\dag, H] | \mu\rangle   \langle \mu | [H,A] | \nu\rangle   (1+e^{-\beta \omega})
\delta(E_\mu-E_\nu -\omega)\nonumber \\
&=&\frac{1}{Z} \sum_{\mu,\nu} e^{-\beta E_\nu} \langle \nu | (E_\mu-E_\nu) A^\dag | \mu\rangle   \langle \mu |(E_\mu-E_\nu) A| \nu\rangle   (1+e^{-\beta \omega})
\delta(E_\mu-E_\nu -\omega)\nonumber \\
&=&\frac{\omega^2}{Z} \sum_{\mu,\nu} e^{-\beta E_\nu} \langle \nu | A^\dag | \mu\rangle   \langle \mu | A| \nu\rangle   (1+e^{-\beta \omega})
\delta(E_\mu-E_\nu -\omega)\nonumber
\\&=& \omega^2 Q_{A^\dag,A}(\omega).
\end{eqnarray}
The successive use of this identity
 $\omega^2 Q_{A^\dag, A}(\omega) = Q_{[H,A]^\dag, [H,A]}(\omega)$  and the definition
 $C_A^k := [H, [H, \cdots [H, A] \cdots ] ]$  given in Theorem 1
 give 
 $$
\omega^{2k} Q_{A^\dag, A}(\omega) = \omega^{2k-2} Q_{[H,A]^\dag, [H,A]}(\omega)
= \omega^{2k-4} Q_{[H, [H,A]]^\dag, [H,[H,A]]}(\omega) = 
\cdots = Q_{{C_A^k}^\dag, C_A^k}(\omega).
 $$
 This completes the proof.
$\Box$

Let $n$ be a nonnegative even integer, and   define a function $f_n: \mathbb R \to \mathbb R$, $g_n: \mathbb R \to \mathbb R$  and 
$h_n: (-1,1) \to \mathbb R$ by
\begin{equation}
f_n(x) := f(x) -\sum_{m=0} ^n \frac{f^{(m)}(0)}{m!} x^m, \ \ g_n(x) := g(x) -\sum_{m=0} ^n \frac{g^{(m)}(0)}{m!} x^m
, \ \ h_n(x) := h(x) -\sum_{m=0} ^n \frac{h^{(m)}(0)}{m!} x^m.
\end{equation}

\begin{figure}[H]
\includegraphics[width=50mm, angle=90
]{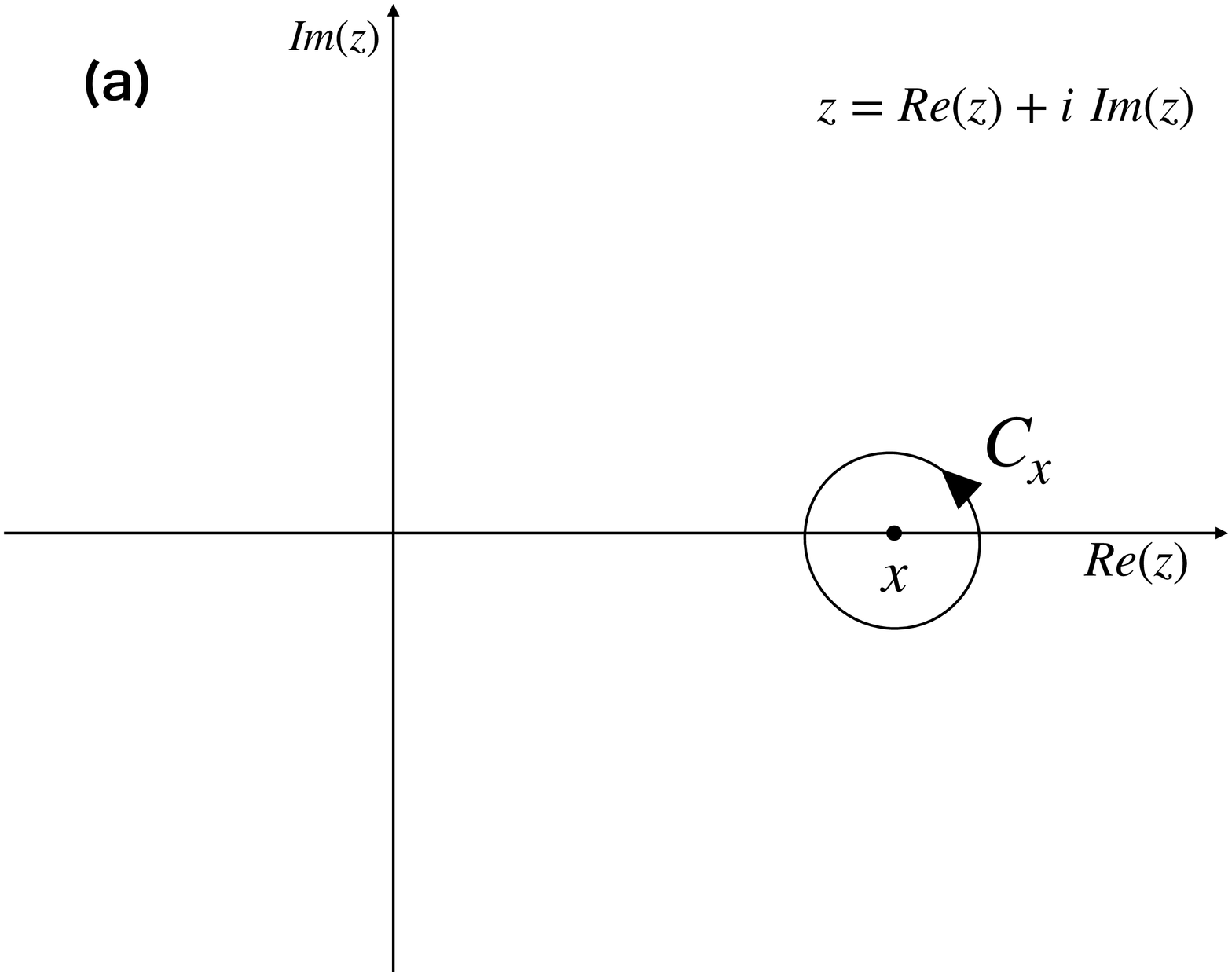}
\includegraphics[width=50mm, angle=90
]{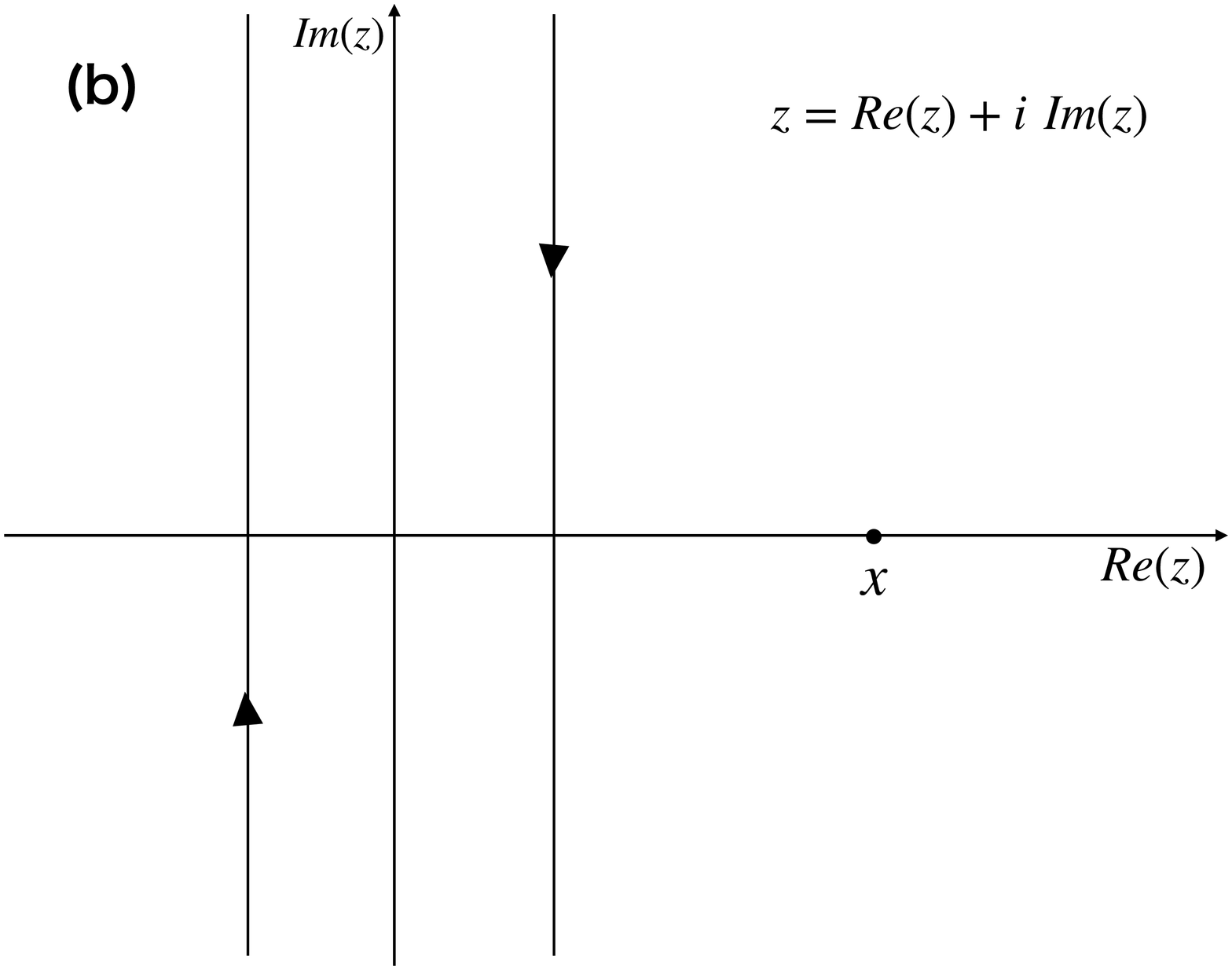}
\caption{{\bf (a) }:The contour for $f^{(n)}(x)$ and $g^{(n)}(x)$,
{\bf (b)}: A deformed contour.}\label{fig}
\end{figure}

\begin{figure}[H]
\includegraphics[width=50mm, angle=90
]{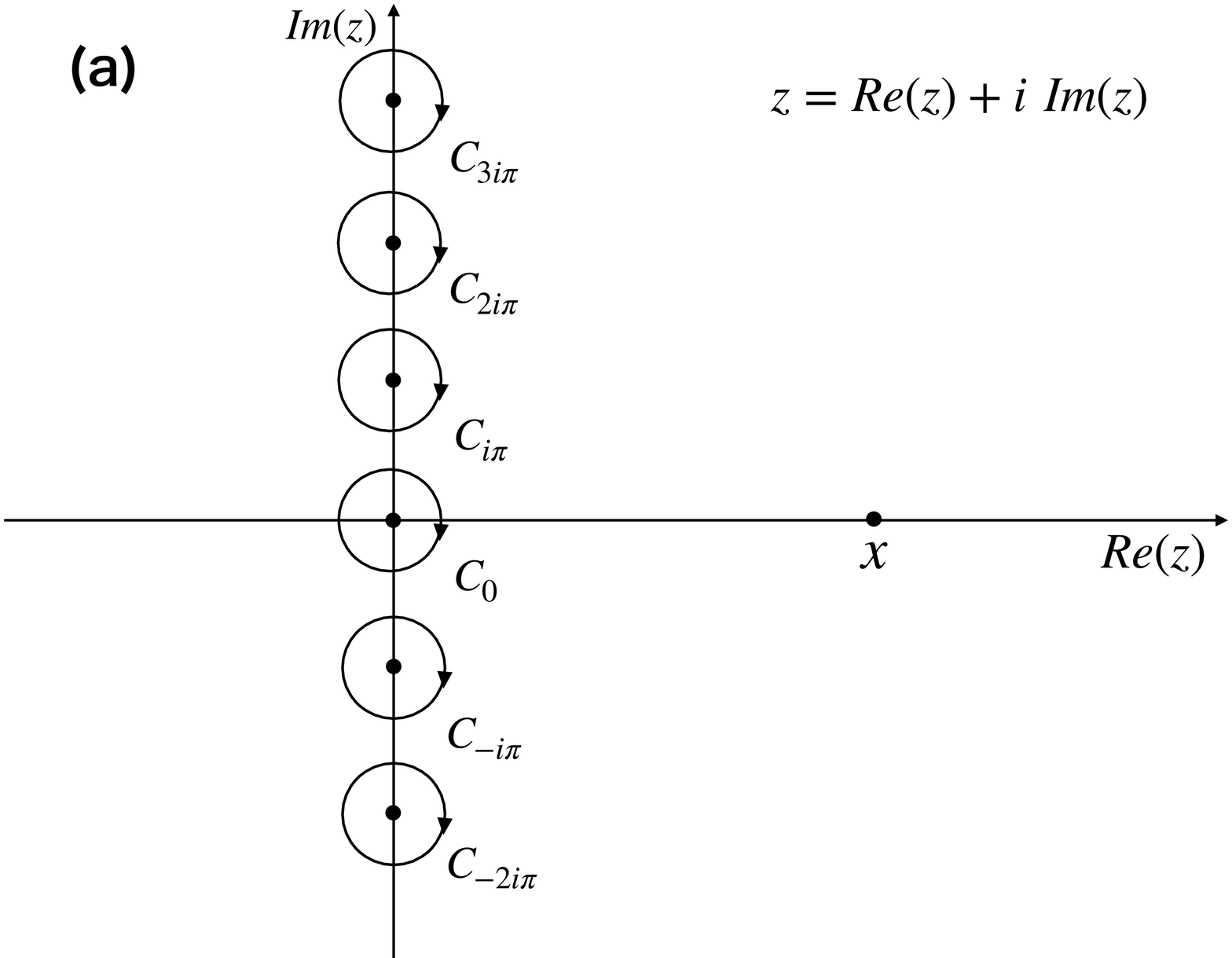}
\includegraphics[width=50mm, angle=90
]{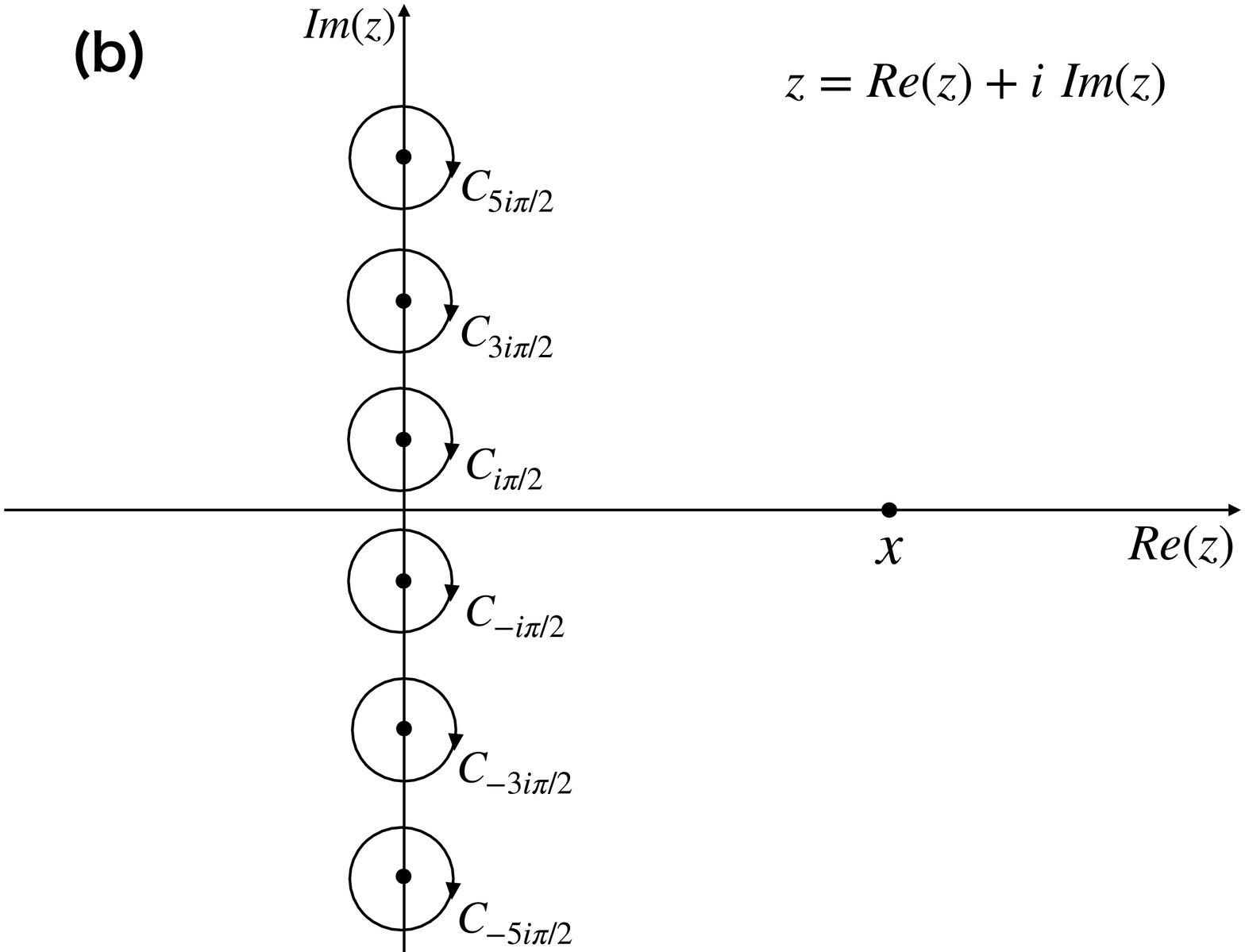}
\caption{{\bf (a)}: Contours for $f^{(n)}(x)$, 
{\bf  (b)}: Contours  for $g^{(n)}(x)$.}\label{fig2}
\end{figure}

\paragraph{Lemma 6}{\it  \label{ML} For any $x \in \mathbb R$ and for any nonnegative even integer $n$,  $f_n(x) \leq 0$ and  $g_n(x) \geq 0$
 for an odd $n/2$, and   $f_n(x) \geq 0$ and  $g_n(x) \leq 0$ for an even $n/2$. 
 For any $x \in (-1,1)$ and for any nonnegative even integer $n$, $h_n(x) \leq 0.$
\\

\noindent
Proof.}   
First, we prove the sign definiteness  of the function $f_n(x)$. Since $f_n(x)$ is an even function, it is sufficient to show the definiteness of
$f_n(x)$ for  $x \geq 0.$ 
 For $x \geq 0$ and for $n=0$,
$$
f_0(x)=f(x) -f(0)= \frac{x-\tanh x}{\tanh x} \geq 0,
$$
since $(x-\tanh x)' = 1-1/ \cosh^2 x \geq 0.$
For a positive even integer $n$, 
$n$-th derivative of the function $f$ is represented in the following contour integral  around $x$ 
depicted in Figure \ref{fig} {\bf (a)}
\begin{equation}
f^{(n)}(x) =n! \ointctrclockwise _{C_x} \frac{dz}{2\pi i} \frac{f(z)}{(z-x)^{n+1}}
=n! \sum_{k=-\infty}^\infty \ointclockwise _{C_{i\pi k}} \frac{dz}{2\pi i} \frac{z \cosh z}{(z-x)^{n+1} \sinh z  }. 
\label{n-th der of f}
\end{equation}
Note that the contour depicted in Figure \ref{fig} {\bf (a)} 
can be deformed into that depicted in Figure \ref{fig} {\bf (b)}. 
Thus, the contour integral (\ref{n-th der of f}) 
is rewritten into that along other contours depicted in Figure \ref{fig2} {\bf (a)} .
The Cauchy formula gives 
\begin{eqnarray}
f^{(n)}(x) &=& n! \sum_{k=-\infty}^\infty \ointclockwise _{C_{i\pi k}} \frac{dz}{2\pi i} \frac{z \cosh z}{(z-x)^{n+1} (z-ik\pi)(\sinh z)'  }\nonumber \\
&=& n! \sum_{k=-\infty}^\infty \ointclockwise _{C_{i\pi k}} \frac{dz}{2\pi i} \frac{i \pi k }{(i \pi k-x)^{n+1} (z-i\pi k) }\nonumber \\
&=& -\frac{n!}{(i\pi)^n} \sum_{k=1}^\infty \Big[\frac{ k }{(k+ix/\pi)^{n+1} }+{\rm c.c.} \Big] \nonumber \\
&=& -\frac{n!}{(i\pi)^n} [-i x/\pi \zeta(n+1, i x/\pi)  +  \zeta(n, i x/\pi) +{\rm  c.c.} ],
\end{eqnarray}
where $\zeta(s,z)$ is the Hurwitz zeta function defined by
$$
\zeta(s,z) := \sum_{k=1}^\infty \frac{ 1 }{(k+z)^{s} },
$$
for $Re s > 1$ and   $z \notin \mathbb Z^-$, where $\mathbb Z^-$ is the set of negative integers. Note that 
\begin{eqnarray}
f_n^{(n)}(x) &=&
f^{(n)} (x) -f^{(n)} (0)  \nonumber \\&=&   -\frac{n!}{(i\pi)^n} \sum_{k=1}^\infty \Big[\frac{ k }{(k+ix/\pi)^{n+1} }+{\rm c.c.} -\frac{2k}{k^{n+1}}\Big]  \nonumber \\
&=& (-1)^{n/2} \ \frac{2n!}{\pi^n} \sum_{k=1}^\infty k^{-n}[1- r_k(x)^{-n-1} \cos (n+1) \theta_k(x) ], 
\label{f^n}
\end{eqnarray}
where $r_k(x) \exp i \theta_k(x) := 1+ix/(\pi k)$. Note that
$$r_k(x) = \sqrt{1+x^2/(\pi k)^2} \geq 1,$$
which implies 
$$1- r_k(x)^{-n-1} \cos (n+1) \theta_k(x)  \geq 0.$$ 
Therefore,  the expression (\ref{f^n}) implies  that for any $x>0$,
$f_n^{(n)} (x) \geq 0$  for even  $n/2$,  and $f_n^{(n)} (x) \leq 0$
for odd $n/2$.  Since  differential coefficients of $f_n$ at 
the origin vanish
$$f_n(0)=0= f_n'(0)=f_n''(0) = \cdots = f_n^{(n-1)}(0).$$
Therefore, $f_n(x) \leq 0$ for an odd $n/2$, and  $f_n(x) \geq 0$ for an even $n/2$.

Next, we prove the sign definiteness  of the function $g_n(x)$. 
Since $g_n(x)$ is an even function, it is sufficient to show the definiteness of
$g_n(x)$ for  $x \geq 0.$ 
 Let $x$  be a nonnegative number and $n$ be a nonnegative even integer. 
The $n$-th order derivative of the function $g$ is represented in the following contour integral  around $x$ depicted in Figure \ref{fig} {\bf (a)} 
\begin{equation}
g^{(n)}(x) =n! \ointctrclockwise _{C_x} \frac{dz}{2\pi i} \frac{g(z)}{(z-x)^{n+1}}
=n! \sum_{k=-\infty}^\infty \ointclockwise _{C_{i\pi (k+1/2)}} \frac{dz}{2\pi i} \frac{ \sinh z}{(z-x)^{n+1} z\cosh z  }.
\label{n-th der of g}
\end{equation}
Note that the contour depicted in Figure \ref{fig} {\bf (a)} 
can be deformed into that depicted in Figure \ref{fig} {\bf (b)}. 
Thus, the contour integral (\ref{n-th der of g}) 
is rewritten into that along other contours depicted in Figure \ref{fig2} {\bf (b)} .
As in the calculation  for $f_n^{(n)}(x)$, $g_n^{(n)}(x)$ can be obtained as
  \begin{eqnarray}
g_n^{(n)}(x) &=&
g^{(n)} (x) -g^{(n)} (0)  \nonumber \\&=&  
 -\frac{n!}{(i\pi)^{n+2}} \sum_{k=1}^\infty \Big[\frac{ 1 }{(k-1/2+ix/\pi)^{n+1} (k-1/2)}+{\rm c.c.} -\frac{2}{(k-1/2)^{n+2}}\Big]  \nonumber \\
&=& (-1)^{n/2+1} \ \frac{2n!}{\pi^{n+2}} \sum_{k=1}^\infty (k-1/2)^{-n-2}[1- s_k(x)^{-n-1} \cos (n+1) \phi_k(x) ], 
\label{g^n}
\end{eqnarray}
where $s_k(x) \exp i \phi_k(x) := 1+ix/[\pi (k-1/2)]$. This implies that
$g_n^{(n)}(x) \geq 0$ for an odd $n/2$, and $g_n^{(n)}(x) \leq 0$ for an even $n/2$.
 Since  differential coefficients of $g_n$ at 
the origin vanish
$$g_n(0)=0= g_n'(0)=g_n''(0) = \cdots = g_n^{(n-1)}(0).$$
Therefore, $g_n(x) \geq 0$ for an odd $n/2$, and  $f_n(x) \leq 0$ for an even $n/2$.

\begin{figure}[H]
\includegraphics[width=35mm, angle=90
]{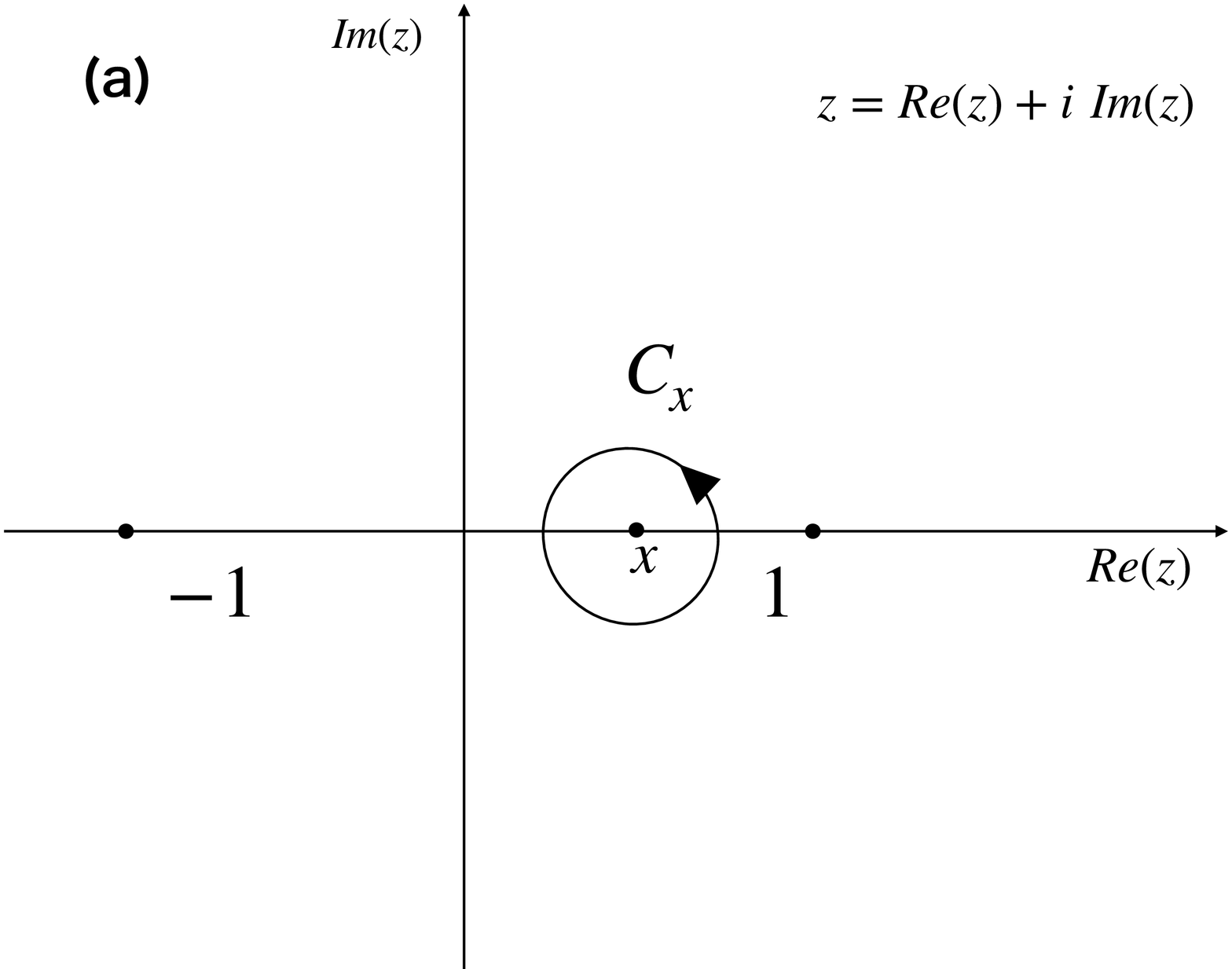}
\includegraphics[width=35mm, angle=90
]{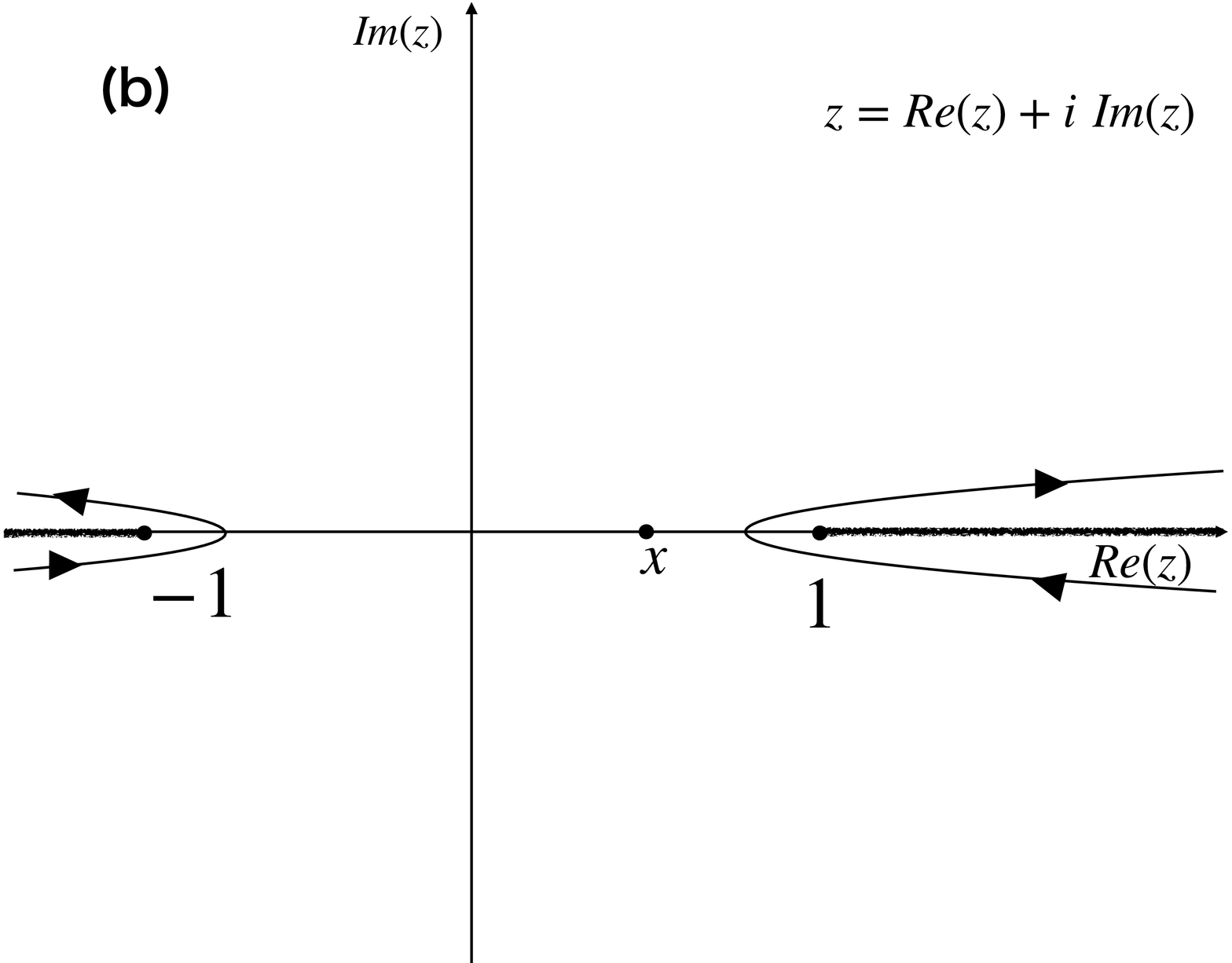}
\includegraphics[width=35mm, angle=90
]{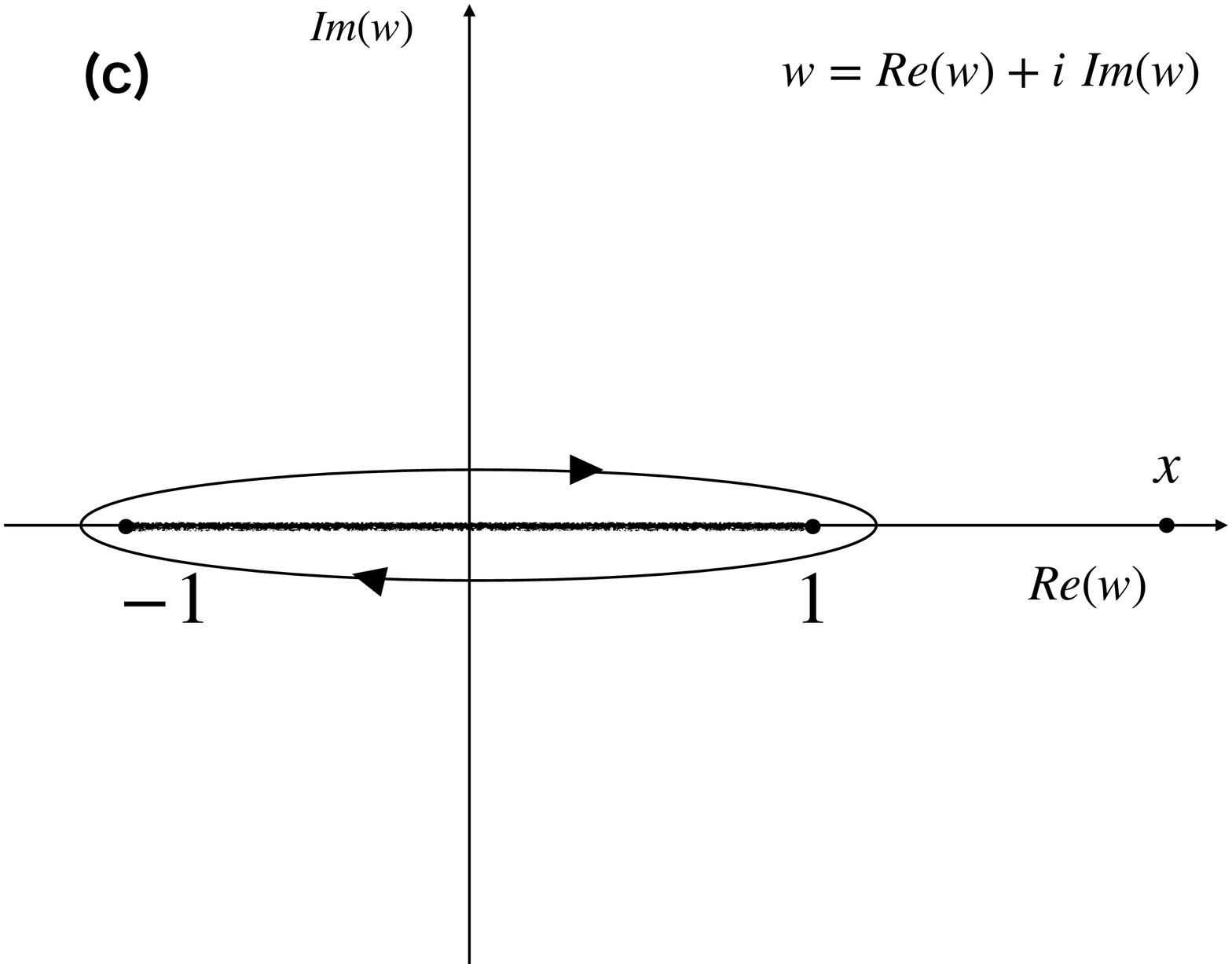}
\caption{{\bf (a)}: Contour for $h^{(n)}(x)$, 
{\bf  (b)}: Deformed contour,  {\bf  (c)}:  integration over $w=1/z$.}
\label{fig3}
\end{figure}

Finally, we prove the sign definiteness  of the function $h_n(x)$. Since $h_n(x)$ is an even function,
 it is sufficient to show the definiteness of
$h_n(x)$ for  $x \in [0,1).$ 
The $n$-th order derivative of the function $h(x)$ can be represented in terms of the following  contour  integral around $x$ depicted in Figure \ref{fig3} {\bf (a)}
\begin{equation}
h^{(n)}(x) =n! \ointctrclockwise _{C_x} \frac{dz}{2\pi i} \frac{h(z)}{(z-x)^{n+1}}
=n!  \ointctrclockwise _{C_x} \frac{dz}{2\pi i} \frac{ z}{(z-x)^{n+1}  \log (1+z)/(1-z)  }.
\end{equation}
Note that the logarithmic function in the integrand has a branch cut $Re z \leq -1, \ Re z \geq 1,$  on the real axis $Im z=0$, as depicted in Figure \ref{fig3} {\bf (b)}.
Rewrite  this integration in terms of  $w=1/z$
\begin{eqnarray}
h^{(n)}(x)
&=&-n!\ointclockwise_{C_{1/x}}\frac{dw}{2\pi i w^2} \frac{ w^{-1}}{(w^{-1}-x)^{n+1}  \log (1+w^{-1})/(1-w^{-1})  }\nonumber  \\
&=&-n! \ointclockwise_{C_{1/x}}\frac{dw}{2\pi i} \frac{ w^{n-2}}{(1-xw)^{n+1}  \log (w+1)/(w-1)  }.
\end{eqnarray}
Note that the logarithmic function in the integrand has a branch cut $-1 \leq  Re w \leq 1,$ on the real axis $Im w=0$,as depicted in Figure \ref{fig3} {\bf (c)}.
This implies 
\begin{eqnarray}
h^{(n)}(x)
&=&-n! \int_{-1} ^1\frac{dw}{2\pi i} \frac{ w^{n-2}}{(1-xw)^{n+1}  }\Big[\frac{1}{ \log (1+w)/(1-w) - \pi i} -\frac{1}{ \log (1+w)/(1-w)+\pi i} \Big]\nonumber \\
&=&-n! \int_{-1} ^1 dw \frac{ w^{n-2}}{(1-xw)^{n+1} [ |\log (1+w)/(1-w) |^2+ \pi^2]} \leq 0,
\end{eqnarray}
for any $x \in (-1,1).$  This fact and 
 $h_n^{(m)}(0) =0$ for any integer $ m \in [0,n] $ imply $h_n(x) \leq 0$ for any $x \in [0,1)$. 
 This completes the proof. $\Box$

\paragraph{ Proof of Theorem 1.}

For a bounded linear operator $A$,
\begin{eqnarray}
&&\frac{1}{2} \langle \{A^\dag, A\} \rangle -(A^\dag,A) =  \int_{-\infty} ^\infty d\omega 
 \Big(\frac{1}{2}-\frac{1}{\beta\omega} \tanh \frac{\beta \omega}{2} \Big) Q_{A^\dag,A} (\omega)
\nonumber \\
 &&= \int_{-\infty} ^\infty  \frac{d\omega}{\beta \omega} \tanh \frac{\beta \omega}{2}
\Big(\frac{\beta \omega}{2} \coth \frac{\beta \omega}{2} - 1 \Big) Q_{A^\dag,A} (\omega).
\end{eqnarray}
For  a positive odd $n/2$, 
Lemma 6,  Lemma 3, 
Lemma 5  and $Q_{A^\dag,A} (\omega) \geq 0$ imply
\begin{eqnarray}
&&\frac{1}{2} \langle \{A^\dag, A\} \rangle -(A^\dag,A)\leq\int_{-\infty} ^\infty  \frac{d\omega}{\beta\omega} \tanh \frac{\beta \omega}{2}
\sum_{k=2} ^{n} \frac{f^{(k)} (0)}{k!} \Big( \frac{\beta \omega}{2}\Big)^kQ_{A^\dag,A} (\omega)\nonumber \\
&&=\int_{-\infty} ^\infty  \frac{d\omega}{2}   \omega \tanh \frac{\beta \omega}{2}
\sum_{k=1} ^{n/2} \frac{f^{(2k)} (0)}{(2k)!} \Big( \frac{\beta}{2}\Big)^{2k-1}\omega^{2k-2}Q_{A^\dag,A} (\omega)
\nonumber \\
&&=\int_{-\infty} ^\infty  \frac{d\omega}{2}\omega \tanh \frac{\beta \omega}{2}
\sum_{k=1} ^{n/2} \frac{f^{(2k)} (0)}{(2k)!} \Big( \frac{\beta}{2}\Big)^{2k-1}Q_{{C_A^{k-1}}^\dag,C_A^{k-1}} (\omega)
\nonumber \\
&&=\sum_{k=1} ^{n/2} \frac{f^{(2k)} (0)}{2(2k)!} \Big( \frac{\beta}{2}\Big)^{2k-1} \langle [{C_A^{k-1}}^\dag,[H,C_A^{k-1}]]   \rangle. 
\end{eqnarray}
Since $(n+ 2)/2$ is even, 
Lemma 6  and $Q_{A^\dag,A} (\omega) \geq 0$ imply
\begin{eqnarray}
&&\frac{1}{2} \langle \{A^\dag, A\} \rangle -(A^\dag,A)\geq\int_{-\infty} ^\infty  \frac{d\omega}{\beta\omega} \tanh \frac{\beta \omega}{2}
\sum_{k=2} ^{n+2} \frac{f^{(k)} (0)}{k!} \Big( \frac{\beta \omega}{2}\Big)^kQ_{A^\dag,A} (\omega)\nonumber \\
&&=\int_{-\infty} ^\infty  \frac{d\omega}{2}\omega \tanh \frac{\beta \omega}{2}
\sum_{k=1} ^{n/2+1} \frac{f^{(2k)} (0)}{(2k)!} \Big( \frac{\beta }{2}\Big)^{2k-1} \omega^{2k-2}Q_{A^\dag,A} (\omega)
\nonumber \\
&&=\int_{-\infty} ^\infty  \frac{d\omega}{2}\omega \tanh \frac{\beta \omega}{2}
\sum_{k=1} ^{n/2+1} \frac{f^{(2k)} (0)}{(2k)!} \Big( \frac{\beta}{2}\Big)^{2k-1}Q_{{C_A^{k-1}}^\dag,C_A^{k-1}} (\omega)
\nonumber \\
&&=\sum_{k=1} ^{n/2+1} \frac{f^{(2k)} (0)}{2(2k)!} \Big( \frac{\beta}{2}\Big)^{2k-1} \langle [{C_A^{k-1}}^\dag,[H,C_A^{k-1}]]   \rangle
\end{eqnarray}
These and $[H,C_A^{k-1}] = C_A^{k}$
complete the proof of Theorem 1. $\Box$\\

\paragraph{ Proof of Theorem 2.}

For a bounded linear operator $A$,
\begin{eqnarray}
&&(A^\dag,A) =  \int_{-\infty} ^\infty d\omega 
\frac{1}{\beta\omega} \tanh \frac{\beta \omega}{2} Q_{A^\dag,A} (\omega)
 = \int_{-\infty} ^\infty  \frac{d\omega}{2} g\Big( \frac{\beta \omega}{2} \Big) Q_{A^\dag,A} (\omega).
\end{eqnarray}
For  a nonnegative even $n/2$, $(n+2)/2$ is an odd integer,
Lemma 6,  Lemma 3, 
Lemma 5  and $Q_{A^\dag,A} (\omega) \geq 0$ imply
\begin{eqnarray}
&&(A^\dag,A)\geq\int_{-\infty} ^\infty  \frac{d\omega}{2}
\sum_{k=0} ^{n+2} \frac{g^{(k)} (0)}{k!} \Big( \frac{\beta \omega}{2}\Big)^kQ_{A^\dag,A} (\omega)
=\int_{-\infty} ^\infty  \frac{d\omega}{2}
\sum_{k=0} ^{n/2+1} \frac{g^{(2k)} (0)}{(2k)!} \Big( \frac{\beta \omega}{2}\Big)^{2k}Q_{A^\dag,A} (\omega)
\nonumber \\
&&=\int_{-\infty} ^\infty  \frac{d\omega}{2} 
\sum_{k=0} ^{n/2+1} \frac{g^{(2k)} (0)}{(2k)!} \Big( \frac{\beta}{2}\Big)^{2k}Q_{{C_A^k}^\dag,C_A^k} (\omega)
=\sum_{k=0} ^{n/2+1} \frac{g^{(2k)} (0)}{2(2k)!} \Big( \frac{\beta}{2}\Big)^{2k} \langle \{ {C_A^k}^\dag,C_A^k \} \rangle. 
\end{eqnarray}
Since $n/2$ is even, 
Lemma 6  and $Q_{A^\dag,A} (\omega) \geq 0$ imply
\begin{eqnarray}
&&(A^\dag,A)\leq\int_{-\infty} ^\infty  \frac{d\omega}{2}
\sum_{k=0} ^{n} \frac{g^{(k)} (0)}{k!} \Big( \frac{\beta \omega}{2}\Big)^kQ_{A^\dag,A} (\omega)
=\int_{-\infty} ^\infty  \frac{d\omega}{2}
\sum_{k=1} ^{n/2} \frac{g^{(2k)} (0)}{(2k)!} \Big( \frac{\beta \omega}{2}\Big)^{2k}Q_{A^\dag,A} (\omega)
\nonumber \\
&&=\int_{-\infty} ^\infty  \frac{d\omega}{2} 
\sum_{k=0} ^{n/2} \frac{g^{(2k)} (0)}{(2k)!} \Big( \frac{\beta}{2}\Big)^{2k}Q_{{C_A^k}^\dag,C_A^k} (\omega)
=\sum_{k=0} ^{n/2} \frac{g^{(2k)} (0)}{2(2k)!} \Big( \frac{\beta}{2}\Big)^{2k} \langle \{{C_A^k}^\dag,C_A^k \} \rangle. 
\end{eqnarray}
This completes the proof of Theorem 2. $\Box$\\

\paragraph{ Proof of Theorem 3.}

For a bounded linear operator $A$,
\begin{eqnarray}
&&\langle [A^\dag,[H,A]]\rangle=  \int_{-\infty} ^\infty d\omega 
\omega \tanh \frac{\beta \omega}{2} Q_{A^\dag,A} (\omega)
 = \int_{-\infty} ^\infty  d\omega\frac{\beta}{2}  g\Big( \frac{\beta \omega}{2} \Big) \omega^2 Q_{A^\dag,A} (\omega).
\end{eqnarray}
For  a nonnegative even $n/2$, $(n+2)/2$ is an odd integer,
Lemma 6,  Lemma 3, 
Lemma 5  and $Q_{A^\dag,A} (\omega) \geq 0$ imply
\begin{eqnarray}
&&\langle [A^\dag,[H,A]]\rangle\geq\int_{-\infty} ^\infty d\omega \frac{\beta}{2}
\sum_{k=0} ^{n+2} \frac{g^{(k)} (0)}{k!} \Big( \frac{\beta \omega}{2}\Big)^k\omega^2 Q_{A^\dag,A} (\omega)
=\int_{-\infty} ^\infty  d\omega
\sum_{k=0} ^{n/2+1} \frac{g^{(2k)} (0)}{(2k)!} \Big( \frac{\beta }{2}\Big)^{2k+1} \omega^{2k+2}Q_{A^\dag,A} (\omega)
\nonumber \\
&&=\int_{-\infty} ^\infty  d\omega
\sum_{k=0} ^{n/2+1} \frac{g^{(2k)} (0)}{(2k)!} \Big( \frac{\beta}{2}\Big)^{2k+1}Q_{{C_A^{k+1}}^\dag,C_A^{k+1}} (\omega)
=\sum_{k=0} ^{n/2+1} \frac{g^{(2k)} (0)}{(2k)!} \Big( \frac{\beta}{2}\Big)^{2k+1} \langle \{ {C_A^{k+1}}^\dag,C_A^{k+1} \} \rangle. 
\end{eqnarray}
Since $n/2$ is even, 
Lemma 6  and $Q_{A^\dag,A} (\omega) \geq 0$ imply
\begin{eqnarray}
&&\langle [A^\dag,[H,A]]\rangle\leq\int_{-\infty} ^\infty  d\omega
\sum_{k=0} ^{n} \frac{g^{(k)} (0)}{k!} \Big( \frac{\beta }{2}\Big)^{k+1}\omega^{k+2}Q_{A^\dag,A} (\omega)
=\int_{-\infty} ^\infty  d\omega
\sum_{k=1} ^{n/2} \frac{g^{(2k)} (0)}{(2k)!} \Big( \frac{\beta}{2}\Big)^{2k+1} \omega^{2k+2}Q_{A^\dag,A} (\omega)
\nonumber \\
&&=\int_{-\infty} ^\infty  d\omega
\sum_{k=0} ^{n/2} \frac{g^{(2k)} (0)}{(2k)!} \Big( \frac{\beta}{2}\Big)^{2k+1}Q_{{C_A^{k+1}}^\dag,C_A^{k+1}} (\omega)
=\sum_{k=0} ^{n/2} \frac{g^{(2k)} (0)}{(2k)!} \Big( \frac{\beta}{2}\Big)^{2k+1} \langle \{{C_A^{k+1}}^\dag,C_A^{k+1} \} \rangle. 
\end{eqnarray}
This completes the proof of Theorem 3. $\Box$\\

\paragraph{ Proof of Theorem 4.}
The spectral representation of the
 Duhamel function for bounded linear operators $A^\dag, A$ is
\begin{equation}
 (A^\dag,A) = \int_{-\infty}^\infty \frac{d\omega}{\beta \omega} \tanh \frac{\beta \omega}{2} Q_{A^\dag, A}(\omega)= \int_{-\infty}^\infty \frac{d\omega}{2} g\Big( \frac{\beta \omega}{2}\Big) Q_{A^\dag, A}(\omega),
\end{equation}
where $g: \mathbb R \to \mathbb R$ defined by (\ref{fgh}).
Define an integration measure 
\begin{equation}
d \mu (\omega) := \frac{d\omega}{2} g\Big( \frac{\beta \omega}{2} \Big)\frac{Q_{A^\dag, A}(\omega)}{ (A^\dag, A)}.
\end{equation}
Note that 
$$
\int_{-\infty} ^\infty d \mu(\omega)  =1.
$$
The Jensen inequality for the 
convex function  $f: \mathbb R \to \mathbb R$ defined by (\ref{fgh}) implies 
\begin{equation}
 f\Big( \frac{\langle [A^\dag,A ]\rangle}{2(A^\dag,A)} \Big)= f\Big( \int_{-\infty} ^\infty d \mu(\omega) \frac{\beta\omega}{2}  \Big)  \leq \int_{-\infty} ^\infty d \mu(\omega)  f\Big(\frac{\beta\omega}{2}\Big)=
 \frac{\langle \{ A^\dag, A\} \rangle}{2 (A^\dag, A)}.
\end{equation}
This inequality gives
\begin{equation} 
  \frac{2(A^\dag,A)}{\langle [A^\dag,A ]\rangle}\tanh  \frac{\langle [A^\dag,A ]\rangle}{2(A^\dag,A)}=  g\Big( \frac{\langle [A^\dag,A ]\rangle}{2(A^\dag,A)} \Big) \geq  \frac{2 (A^\dag, A)}{\langle \{ A^\dag, A\} \rangle},
\end{equation}
then 
\begin{equation} 
\tanh  \frac{\langle [A^\dag,A ]\rangle}{2(A^\dag,A)}\geq  \frac{\langle [A^\dag,A ]\rangle}{\langle \{ A^\dag, A\} \rangle}.
\end{equation}
This inequality can be represented in terms of the function $h : (-1,1) \to \mathbb R$ defined by (\ref{fgh})
\begin{equation}
\frac{(A^\dag, A)}{ \langle \{ A^\dag, A\} \rangle}\leq
 \frac{\frac{\langle [A^\dag,A ]\rangle}{\langle \{ A^\dag, A\} \rangle}}{2 \tanh^{-1}   \frac{\langle [A^\dag,A ]\rangle}{\langle \{ A^\dag, A\} \rangle}}= h\Big(\frac{\langle [A^\dag,A ]\rangle}{\langle \{ A^\dag, A\} \rangle} \Big),
\end{equation}
which is obtained by Roepstorff \cite{R}. Lemma 6  gives an upper bound on the right hand side
$$
h\Big(\frac{\langle [A^\dag,A ]\rangle}{\langle \{ A^\dag, A\} \rangle} \Big) \leq \sum_{k=0}^n \frac{h^{(k)}(0)}{k!} \Big(\frac{\langle [A^\dag,A ]\rangle}{\langle \{ A^\dag, A\} \rangle} \Big)^k.
$$
Therefore, 
$$
\frac{(A^\dag, A)}{ \langle \{ A^\dag, A\} \rangle}\leq \sum_{k=0}^n \frac{h^{(k)}(0)}{k!} \Big(\frac{\langle [A^\dag,A ]\rangle}{\langle \{ A^\dag, A\} \rangle} \Big)^k,
$$ 
is obtained. This  completes the proof of
 Theorem 4. $\Box$

 \section{Applications to the Transverse Field Sherrington-Kirkpatrick  Model}
Here, we study  quantum spin systems with random interactions. Let $N$ be a positive integer and a site index 
$i \ (\leq  N)$ is also a positive integer. 
A sequence of spin operators 
 $(\sigma^{w}_i)_{w=x,y,z, 1\leq i \leq  N}$ on a Hilbert space ${\cal H} :=\bigotimes_{i =1}^N {\cal H}_i$ is
defined by a tensor product of the Pauli matrix $\sigma^w$ acting on ${\cal H}_i \simeq {\mathbb C}^{2}$ and unities.
These operators are self-adjoint and satisfy the commutation relation
$$
 [\sigma_k^y,\sigma_j^z]=2i \delta_{k,j} \sigma_j^x ,\ \ \  \ \ 
[\sigma_k^z,\sigma_j^x]=2i \delta_{k,j} \sigma_j^y ,\ \  \ \ \ [\sigma_k^x,\sigma_j^y]=2i \delta_{k,j} \sigma_j^z ,  
$$
and each spin operator satisfies
$$
(\sigma_j^w)^2 = {\bf 1}.
$$
The Sherrington-Kirkpatrick (SK) model is 
  well-known as a disordered classical  spin  system \cite{SK}.
The transverse field  SK  model 
 is a simple quantum  extension.
Here, we study a magnetization process for a local field in these models.
Consider  the following Hamiltonian with  coupling constants
$b_1, h\in {\mathbb R}$ 
\begin{equation}
H_N( \sigma, b_1, g,  h):=- \frac{1 }{\sqrt{N}}\sum_{1\leq i<j\leq N} g_{i,j}  \sigma_i^z \sigma_j^z
-\sum_{j=1}^N h\sigma_j^z-\sum_{j=1}^N b_1 \sigma_j^x,
\label{hamil}
\end{equation}
where  $g=(g_{i,j})_{1\leq i<j \leq N}$ 
is a sequence of independent standard Gaussian random variables
obeying a probability density function
\begin{equation}
p(g):=  \prod_{1\leq i<j\leq N}
 \frac{1}{\sqrt{2\pi}} e^{-\frac{g_{i,j} ^2}{2}}
\label{distributiong}
\end{equation}
The Hamiltonian is invariant under $\mathbb Z_2$-symmetry $U \sigma_i ^z U^\dag = -\sigma_i^z$
for the discrete unitary transformation $U:= \prod_{1\leq i \leq N} \sigma_i^x$
 for $ h= 0$.  
For a positive $\beta $, the  partition function is defined by
\begin{equation}
Z_N(\beta, b_1,  g,  h) := {\rm Tr} e^{ - \beta H_N( \sigma,b_1,  g,  h )},
\end{equation}
where the trace is taken over the Hilbert space ${\cal H}$.
Here, we define a square root interpolation for the transverse field SK
model, as for the SK model given by Guerra and Talagrand  \cite{G1,T}. 
Let  $z=(z_j)_{1\leq j \leq N} $ be a sequence of independent standard  Gaussian random variables. 
This method gives a variational solution of specific free energy.
Consider the following interpolated  Hamiltonian  with  parameters $s \in [0,1]$, $b(s):=b_1s+b_0(1-s)$
for  $b_0 \in \mathbb R$ and $q \in [0,1]$
\begin{eqnarray}
H(s, \sigma):=
- \sqrt{\frac{s}{N}}\sum_{1\leq i<j\leq N} g_{i,j}  \sigma_i^z \sigma_j^z-\sum_{j=1}^N (\sqrt{q(1-s)} z_j +h)\sigma_j^z
-\sum_{j=1}^N b(s) \sigma_j^x.
\end{eqnarray}
Define an interpolated function $\varphi_N(s)$
\begin{equation}
\varphi_N(s) :=\frac{1}{N} \mathbb E \log {\rm Tr} e^{-\beta H(s, \sigma)}
\end{equation}
where $\mathbb E$ denotes the expectation over all Gaussian random variables $(g_{i,j})_{1 \leq i<j \leq N}$  and $(z_i)_{1\leq i\leq N}$.
Note that $\varphi_N(1)$ is given by
$$
\varphi_N(1) = \frac{1}{N}\mathbb E \log Z_N(\beta, b_1,g,h),
$$
which is proportional to the specific free energy of the transverse field SK model.
Let  $f$ be an arbitrary function 
of a sequence of spin operators $\sigma=(\sigma_i^w)_{ w=x,y,z, 1 \leq i \leq N}$.  
The  expectation of $f$ in the Gibbs state is given by
\begin{equation}
\langle f( \sigma) \rangle_s=\frac{{\rm Tr} f( \sigma)  e^{ - \beta H(s,\sigma)}}{{\rm Tr}  e^{ - \beta H(s,\sigma)}}.
\end{equation}
The derivative of $\varphi_N(s)$ with respect to $s$
is given by
\begin{equation}
\varphi'_N(s)= \frac{\beta}{2N^\frac{3}{2}\sqrt{s}} \sum_{1\leq < j\leq N} \mathbb E g_{i,j}\langle \sigma_i^z \sigma_j^z \rangle_s - \frac{\beta\sqrt{q}}{2N\sqrt{1-s}} \sum_{i=1}^N \mathbb E z_i \langle \sigma_i^z \rangle_s+
\frac{\beta b'(s)}{N} \sum_{i=1}^N \mathbb E\langle \sigma_i^x \rangle_s.
\end{equation}
The identities for the Gaussian random variables $g_{i,j}$ and $z_i$
$$
g_{i,j} p(g, z) = -\frac{\partial p}{\partial g_{i,j}},   \ \ \ z_i p(g, z) =- \frac{\partial p}{\partial z_i}
$$
and the integration by parts imply
\begin{eqnarray}
\varphi'_N(s)&=& \frac{\beta^2}{2N^2} \sum_{1\leq < j\leq N} \mathbb E [(\sigma_i^z \sigma_j^z , \sigma_i^z \sigma_j^z )_s-\langle \sigma_i^z \sigma_j^z \rangle_s^2 ]- \frac{\beta^2q}{2N} \sum_{i=1}^N \mathbb E [(\sigma_i^z  , \sigma_i^z )_s-\langle \sigma_i^z \rangle_s^2 ]+
\frac{\beta b'(s)}{N} \sum_{i=1}^N \mathbb E\langle \sigma_i^x \rangle_s \nonumber \\
&=&\frac{\beta^2(N-1)}{4N}[\mathbb E(\sigma_1^z \sigma_2^z , \sigma_1^z \sigma_2^z )_s
-1 ]- \frac{\beta^2q}{2}[ \mathbb E (\sigma_1^z  , \sigma_1^z )_s-1]+
\beta b'(s) \mathbb E\langle \sigma_1^x \rangle_s \nonumber \\
&+&\frac{\beta^2}{4} (1-q)^2-\frac{\beta^2}{4} \mathbb E \langle (R_{1,2}-q) ^2\rangle_s
, \label{phi'}
\end{eqnarray}
where
 the overlap operator $R_{a,b}$ is defined by 
$$
R_{a,b} := \frac{1}{N} \sum_{i=1}^N \sigma_i^{z,a} \sigma_i^{z,b},
$$
for independent replicated Pauli operators $\sigma_i^{z,a} \ (a= 1, 2,\cdots, n)$ obeying the same Gibbs state with
the replica Hamiltonian
$$
H(s, \sigma^1, \cdots, \sigma^n):= \sum_{a=1}^n H(s, \sigma^a).
$$
This Hamiltonian is invariant under 
 permutation of replica spins. This permutation symmetry is known to be the replica symmetry. 
The order operator $R_{a,b}$ measures the replica symmetry breaking as an order operator. 
In the identity (\ref{phi'}), we use
an upper bound 
$\mathbb{E}(\sigma_1^z \sigma_2^z , \sigma_1^z \sigma_2^z )_s\le 1$
given by the right-hand side of the inequality (\ref{Ex2}),
and  
the lower bound on the Duhamel function $(\sigma_1^z,\sigma_1^z)$  %
$$
-\frac{\beta^2 q}{2}\mathbb{E}[(\sigma_1^z  , \sigma_1^z )_s-1]
\le\frac{\beta^4q}{48}\mathbb{E}\langle \{[H,\sigma_1^z  ]^\dag, [H,\sigma_1^z]\} \rangle_s
=\frac{\beta^4q}{6}b(s)^2,
$$
given by the left-hand side of the inequality (\ref{Ex2}). 
%
Then, we have
\begin{eqnarray}
\varphi'_N(s) &\leq&
\frac{\beta^4q}{48} \mathbb E \langle \{[H,\sigma_1^z  ]^\dag, [H,\sigma_1^z]\} \rangle_s+
\beta b'(s)\mathbb E\langle \sigma_1^x \rangle_s+\frac{\beta^2}{4} (1-q)^2-\frac{\beta^2}{4} \mathbb E \langle (R_{1,2}-q) ^2\rangle_s \nonumber \\
&=& 
\frac{\beta^4q}{6} b(s)^2+\beta b'(s)\mathbb E\langle \sigma_1^x \rangle_s+\frac{\beta^2}{4} (1-q)^2-\frac{\beta^2}{4} \mathbb E \langle (R_{1,2}-q) ^2\rangle_s \nonumber \\
&\leq & \frac{\beta^4q}{6} b(s)^2+\beta b'(s)\tanh \beta b(s)+\frac{\beta^2}{4} (1-q)^2, 
\label{9FB}
\end{eqnarray}
where an upper bound $\tanh \beta b(s) \geq \langle \sigma_1^x \rangle_s$ has been used as shown by Leschke, Manai, Ruder and Warzel \cite{W}. 
%
The  bound on the Duhamel function $(\sigma_1^z  , \sigma_1^z )_s$
can be obtained also from 
the Falk-Bruch inequality \cite{FB}  and our result for $g(z)$ defined by (\ref{fgh}), instead of the simple use of the
inequality (\ref{Ex2}).
For $\Phi(r\tanh r):=\tanh r/r=g(r)$, 
$$
(\sigma_1^z  , \sigma_1^z )_s
\ge
\Phi\biggl(\frac{\langle [\sigma_1^z,[\beta H,\sigma_1^z] ]\rangle_s}{4}\biggr),
$$
 is obtained by Leschke, Manai, Ruder and Warzel 
as  a corollary  of the Falk-Bruch inequality \cite{W}. 
The equality and the inequality
$\langle [\sigma_1^z,[\beta H,\sigma_1^z] ]\rangle_s=4\beta b(s)\langle \sigma_1^x \rangle_s\le4\beta b(s)\tanh\beta b(s)$, the monotonicity and the convexity of the function
$\Phi$, 
the definition of $\Phi$ and Lemma 6 give  the lower bound on $(\sigma_1^z  , \sigma_1^z )_s$ 
$$
(\sigma_1^z  , \sigma_1^z )_s\ge\Phi(\beta b(s)\langle \sigma_1^x \rangle_s)
\ge \Phi(\beta b(s)\tanh\beta b(s))=g(\beta b(s))
\ge1-\frac{\beta^2}{3}b(s)^2,
$$
which gives  the same upper bound 
(\ref{9FB}). 
The advantage of using
the new inequality (\ref{Ex2}) in the present case is that 
the lower bound  on the Duhamel function is easily expressed in terms of the known simple  function 
of $\beta$ and $b(s)$ with far fewer calculations.  
Although a relation between the Falk-Bruch inequality and the new inequality (\ref{Ex2}) can be understood
in the present specific case, 
it is difficult to clarify that in general case. 

Integration %
of the inequality ({\ref{9FB}}) over %
$s\in[0,1]$ gives 
\begin{eqnarray}
\varphi_N(1) &\leq& \varphi_N(0) + \int_0 ^1 ds [\frac{\beta^4q}{6} b(s)^2+\beta b'(s)\tanh \beta b(s)+\frac{\beta^2}{4} (1-q)^2] \nonumber \\
&=&  \varphi_N(0) + \frac{\beta^4q}{18} (b_0^2+b_0b_1+b_1^2) +  \log \frac{\cosh \beta b_1}{\cosh \beta b_0}+ \frac{\beta^2}{4} (1-q)^2=:\Phi(q,b_0).
\label{TFSK}
\end{eqnarray}
The model at $s=0$ becomes independent spin model, and therefore 
\begin{equation}
\varphi_N(0) = \mathbb E \log {\rm Tr} \exp \beta[ (\sqrt{q} z+h) \sigma^z + b_0 \sigma^x]
=  \mathbb E \log 2 \cosh \beta \sqrt{(\sqrt{q} z +h)^2+ b_0^2}.
\end{equation}
A variational solution with the best bound
is obtained by minimizing the right hand side in (\ref{TFSK}). The minimizer $(q, b_0)$ should satisfy
\begin{eqnarray}
0&=&\frac{\partial}{\partial q} \Phi(q,b_0)=\mathbb E\frac{\beta z( \sqrt{q}z +h)  \tanh \beta \sqrt{(\sqrt{q} z+h)^2+ b_0^2}}{2\sqrt{q}\sqrt{(\sqrt{q} z+h)^2+ b_0^2}} 
+\frac{\beta^4}{18} (b_0^2+b_0b_1+b_1^2) - \frac{\beta^2}{2} (1-q) \nonumber
 \\
 &=& \mathbb E\frac{\beta^2 (\sqrt{q}z +h)^2}{2[(\sqrt{q} z+h)^2+ b_0^2] \cosh^{2} \beta \sqrt{(\sqrt{q} z+h)^2+ b_0^2}} +\mathbb E\frac{  \beta b_0^2 \tanh  \beta \sqrt{(\sqrt{q} z+h)^2+ b_0^2}}{2[(\sqrt{q} z+h)^2+ b_0^2]^\frac{3}{2}} 
\nonumber \\
&+&\frac{\beta^4}{18} (b_0^2+b_0b_1+b_1^2) - \frac{\beta^2}{2} (1-q), \label{Phi_q}\\
0&=& \frac{\partial}{\partial b_0} \Phi(q,b_0) =\mathbb E\frac{\beta b_0 \tanh \beta \sqrt{(\sqrt{q} z+h)^2+ b_0^2}}{\sqrt{(\sqrt{q} z+h)^2+ b_0^2}} -\beta \tanh \beta b_0
+\frac{\beta^4 q}{18} (2b_0+b_1), \label{Phi_b0}
\end{eqnarray}
where the following integration by parts has been used to obtain the equation (\ref{Phi_q})
$$
\mathbb E\frac{\beta z( \sqrt{q}z +h)  \tanh \beta \sqrt{(\sqrt{q} z+h)^2+ b_0^2}}{2\sqrt{q}\sqrt{(\sqrt{q} z+h)^2+ b_0^2}} 
=\mathbb E\frac{\beta }{\sqrt{q}}\frac{\partial }{\partial z}\frac{( \sqrt{q}z +h)  \tanh \beta \sqrt{(\sqrt{q} z+h)^2+ b_0^2}}{2\sqrt{(\sqrt{q} z+h)^2+ b_0^2}}. 
$$
This minimizer $(q,b_0)$ gives the best bound on $\varphi_N(1)$ as a variational solution
\begin{equation}
\varphi_N(1) \leq  \mathbb E \log 2 \cosh \beta \sqrt{(\sqrt{q} z +h)^2+ b_0^2}+ \frac{\beta^4q}{18} (b_0^2+b_0b_1+b_1^2) +  \log \frac{\cosh \beta b_1}{\cosh \beta b_0}+ \frac{\beta^2}{4} (1-q)^2.
\label{phi1}
\end{equation}
Note that the equation
(\ref{Phi_b0}) 
 has a solution $b_0 =0$ in the classical limit $b_1\to0$. In this case the equation (\ref{Phi_q})
  becomes 
 \begin{equation}
q= \mathbb E \tanh^2 \beta( \sqrt{q} z +h).  \label{classical}
\end{equation}
 Then,  the solution (\ref{phi1}) is identical to the SK solution \cite{SK} 
in the classical limit $b_1\to0$. In the classical case $b_1=0$,
 it is conjectured  that the replica symmetry is preserved  with 
 $$
 \lim_{N\to \infty}\mathbb E \langle (R_{1,2}-q)^2\rangle_1 =0, 
$$  and the SK solution of the specific free energy is exact
 for
$$
\mathbb E \frac{\beta^2}{\cosh^4 \beta (\sqrt{q}z +h)} \leq 1,
$$
whose boundary is called  the Almeida-Thouless line \cite{AT}.
Recently, Chen has proven rigorously that the SK solution is exact for independent centered Gaussian random external fields,
instead of the uniform field $h$
\cite{WK-C}.   For the uniform field $h\neq 0$, it still remains a conjecture.

Consider a simple case $h=0$, where the model has the $\mathbb Z_2$ symmetry.
If the replica symmetric solution  $q=0$ is assumed in this case, the equation (\ref{Phi_q}) becomes
\begin{equation}
0= \tanh \beta b_0 -\beta b_0 + \frac{\beta^3}{9}b_0(b_0^2+b_0b_1+b_1^2), \label{eq-b0}
 \end{equation}
 which fixes $b_0$, and the equation (\ref{Phi_b0}) is valid for any $b_0$. 
 Therefore, the $\mathbb Z_2$ and  replica symmetric variational solution 
 of the specific free energy
 is given by
\begin{equation}
-\varphi_N(1)/\beta \geq -\frac{1}{\beta}\log 2\cosh \beta b_1- \frac{\beta}{4},
\label{h=0}
\end{equation}
under the assumption $q=0$ for $h=0$. 
 This lower bound can be compared to results obtained in other literatures.   
Leschke, Rothlauf, Ruder and Spitzer evaluate the specific free energy in %
a different rigorous %
method based on the annealed free energy
 \cite{LRRS}.
They first give %
a simple estimate of its lower
bound %
in the high temperature region $\beta < 1$ \cite{LRRS}.
This lower bound is exactly the same as the right hand side of the inequality %
(\ref{h=0}) %
in the infinite volume limit. 
In addition, they obtain a corrected estimate in a high temperature expansion \cite{LRRS}.
Although this estimate might be better, 
the correction is quite small. 
The specific free energy $%
f_{\rm st}(\beta,b_1,h) $ obtained by the replica trick with
the static approximation \cite{KK} %
is
\begin{equation}
-\varphi_N(1)/\beta \simeq %
f_{\rm st}(\beta,b_1,h=0)= -\frac{1}{\beta}\log 2\cosh \beta b_1,
\label{st}
\end{equation}
which violates a rigorous upper bound 
\begin{equation}
-\varphi_N(1)/\beta \leq -\frac{1}{\beta}\log 2\cosh \beta b_1- \frac{\beta}{8}\Big(\frac{1}{\cosh^2 \beta b_1 }+\frac{\tanh \beta b_1}{\beta b_1}\Big) ,
\end{equation}
given by Leschke, Rothlauf, Ruder and Spitzer\cite{LRRS}. 
%
For a strong field $h\gg 1$, however, 
the approximate specific free energy
$$
f_{\rm st}(\beta,b_1,h)\simeq -\frac{1}{\beta}\mathbb E \log 2 \cosh \beta \sqrt{(\sqrt{q} z +h)^2+ b_1^2},
$$
must be a good approximation, since the following deviation of the strong field limit vanishes 
$$\lim_{h\to\infty}[\beta f_{\rm st}(\beta, b_1, h)+\varphi_N(1)] =0.
$$   
On the other hand, 
the upper bound (\ref{phi1}) and $q\to 1$ 
in this limit  
give an upper bound on the following deviation 
\begin{eqnarray*}
\lim_{h\to\infty}[\varphi_N(1) -\mathbb E \log 2 \cosh \beta \sqrt{(\sqrt{q} z +h)^2+ b_1^2}] \leq
\frac{\beta^4}{18} (b_0^2+b_0b_1+b_1^2) + \log \frac{\cosh \beta b_1}{\cosh \beta b_0} =: D,
\end{eqnarray*}
where $b_0$ is the solution of the equation 
$$
\tanh \beta b_0
-\frac{\beta^3 }{18} (2b_0+b_1)=0,
$$
obtained from (\ref{Phi_b0}). Since the relative deviation vanishes
$$
\lim_{h\to \infty} D/ \varphi_N(1)=0,
$$ 
in the strong field limit, the upper bound on $\varphi_N(1)$ given by the right hand side of (\ref{phi1})
must be a  good approximation for $h\gg1$ as well as the approximate specific free energy
$f_{\rm st}(\beta, b_1,h)$.  
\\

\noindent
{\bf Acknowledgments \ } \\
\\
C.I. is supported by JSPS (21K03393).
 
\end{document}